\documentclass[a4,aps,amsmath,floatfix]{revtex4}
\usepackage{graphicx} \usepackage{amsmath} \usepackage{enumerate} 
\usepackage{braket} \usepackage[caption=false]{subfig} 
\usepackage[colorlinks=true,linkcolor=red]{hyperref} \usepackage{xcolor} 
\usepackage{hyperref} \newcommand{\be}{\begin{equation}} 
\newcommand{\ee}{\end{equation}} \newcommand{\bea}{\begin{eqnarray}} 
\newcommand{\eea}{\end{eqnarray}} \newcommand{\bdm}{\begin{displaymath}} 
\newcommand{\edm}{\end{displaymath}} 
\usepackage{subfig} \usepackage{amsmath} 
\usepackage{enumerate} 
\begin{document}
\large{ 

\large{ \title{Note on Boltzmann's H-Theorem and Detailed Balance 
Dynamics}

\author{Prabodh Shukla} \affiliation{Retired, North-Eastern Hill 
University, Shillong, India.} \date{\today}

\begin{abstract} {\large{ The Boltzmann H-theorem states that entropy of 
an ensemble of thermodynamic states increases until all states become 
equally probable. The evolution randomly picks one of the states in the 
ensemble and brings every other state to have the same probability as 
the selected state. Different realizations of evolution yield a 
distribution of selected states. We study this distribution numerically 
and discuss its relevance for equilibrium fluctuations in thermal as 
well as non thermal systems.}} \end{abstract}

\maketitle

\section{Introduction}

Boltzmann's work provides a microscopic support to 
thermodynamics~\cite{reif}. He postulated that a system in equilibrium 
with a heat reservoir at temperature $T$ may occupy one of its 
accessible states of energy $\epsilon_i$ with the probability,

\bdm p_i=\frac{1}{Z}\exp^{-\frac{\epsilon_i}{k_BT}}, 
Z=\sum_i\exp^{-\frac{\epsilon_i}{k_BT}} \edm

Here $k_B$ is known as Boltzmann's constant, and $Z$ is a normalizing 
factor for the probability. $Z$ is called the partition function and 
plays a central role in connecting statistical mechanics to 
thermodynamic functions like Helmholtz free energy $F$, internal energy 
$E$, and entropy $S$. Each term in the sum $Z$ comprises a product of 
Boltzmann factor and degeneracy of energy $\epsilon_i$. In a 
thermodynamic ally large system, we can replace the discrete sum by an 
integral over continuum of energy states weighted with density of states 
$\Omega(\epsilon_i)$. $\Omega(\epsilon_i)$ rises exponentially rapidly 
with $\epsilon_i$ while the Boltzmann factor decreases exponentially. 
The main contribution comes from a sharp region where the two terms 
balance each other and, Thus,

\bdm Z=\Omega(E)\exp^{-\frac{E}{k_BT}} \mbox{ or }-k_B T \ln{Z}= E - k_B 
T \ln{\Omega} \edm

Thermodynamic relation $F=E - T S$ is recovered by identifying $F=-k_B 
\ln{Z}$ and $S=k_B \ln{\Omega}$. In the following we are concerned with 
an expression for entropy in terms of the probabilities $\{p_i\}$ of 
states accessible to the system. We rewrite,

\bdm S=k_B .\left[\ln{Z} + \frac{1}{k_bT} \sum_i p_i\epsilon_i .\right] 
= \ln{Z} - \sum_i p_i \ln{p_i Z} = -k_b \sum_i p_i \ln{p_i} \mbox{ using 
} \sum_i p_i=1 \edm
 
Following Boltzmann, one considers a differential equation for 
$H=-S/k_B=\sum_ip_i\ln{p_i}$,

\bdm \frac{dH}{dt} = \sum_i \frac{dp_i}{dt}(\ln{p_i}+1) = \sum_i 
 \sum_j \{ p_j W_{i \leftarrow j } - p_i W_{i \rightarrow j} \}
(\ln{p_i}+1) \edm

Here $W_{i \leftarrow j }$ and $W_{i \rightarrow j}$ are respectively 
the transition rates from state $j$ to state $i$ and vice verse. Assuming 
$W_{i \leftarrow j } = W_{i \rightarrow j} = W$, the equation is written 
in a more symmetric form with respect to summation variables $i$ and 
$j$,

\bdm \frac{dH}{dt} = -\frac{1}{2} W \sum_i \sum_j (p_i - p_j) (\ln{p_i}- 
\ln{p_j}) \le 0, \mbox{  or  } \frac{dS}{dt} \ge 0 \edm

This yields Boltzmann's theorem that entropy increases monotonically 
till all states in the ensemble acquire equal probability, say 
$p_i=p_j=p^*$ ~\cite{reif}. The ensemble of states, each with 
probability $p^*$,may be thought to an equilibrium ensemble because it 
is stationary with time. Note that Boltzmann's theorem does not 
determine the value $p^*$. A good deal of intrigue has been associated 
with this theorem since the times of Boltzmann because it apparently 
bestows a direction to the "arrow of time" in contrast to other 
fundamental laws of physics which remain invariant under time reversal. 
We note that key ingredient of the result is the principle of detailed 
balance, i.e. $p_i$ can go to $p_j$ with the same transition rate as 
$p_j$ to $p_i$ at a given time $t$. This is distinct from the principle 
of time reversal invariance.

\section{Numerical Simulations}

It is instructive to verify Boltzmann's theorem by numerical 
simulations. Numerical simulation of a differential equation is 
necessarily based on a finite difference equation. Setting the Boltzmann 
constant $k_B=1$, and also the transition rate $W=1$, the incremental 
change in entropy is,

\be S(t+1)-S(t) = -\sum_i [p_(t+1)\ln{p_i(t+1)}-p_(t)\ln{p_i(t)}] \ee

The above quantity does come to zero after sufficient number of Monte 
Carlo steps but it does not decrease monotonically as predicted by the 
differential equation. We shall return to this point shortly but we 
first look at a quantity that does show monotonic behavior. It is the 
number of probabilities in the set $\{p_i\}$ that acquire equal values. 
This result applies to different initial distributions $\{p_i\}$, not 
just the Boltzmann distribution of energies at a temperature $T$. For 
example, the following considerations are also applicable to entropy in 
information theory ~\cite{shannon}. In the following simulations we take 
a rectangular distribution of $p_i$ in the range [0,1]. We update 
probabilities $\{ p_{i}(t)\}$ in parallel at discrete time steps $t$. A 
key aspect of Boltzmann's update rule is that initial probabilities in 
the set do not change intrinsically but any probability $p_i$ can assume 
a value $p_j$ at next time step $t+1$. Further, $p_j(t)$ may take the 
value $p_i(t)$ with the same probability as $p_i$ to $p_j$. This is the 
detailed balance property which is most important for any statistical 
system to reach equilibrium.

Figure-1 shows how an increasing number of states in an ensemble, having 
different probabilities initially, gradually acquire equal probability 
under detailed balance dynamics. The value of equal probability is the 
probability of a randomly selected initial state in the ensemble. The 
figure consists of two parts, (a) and (b). Fig.1(a) depicts the result 
of simulations of an ensemble of $10^3$ states with initial 
probabilities $\{p_i\}$ drawn from a rectangular distribution over the 
range [0,1]. The data is averaged over $10^3$ runs of simulations. Each 
simulation equalizes the probabilities of all states in the ensemble 
within about 150 iterations of dynamics i.e. 150 Monte Carlo steps of 
parallel updating. We may refer to terminal state of ensemble as a fixed 
state because the probability of each state in the terminal state 
remains unchanged under further updates. Fig.1(b) shows a similar 
simulation as in Fig.1(a) but a slightly different update algorithm. In 
Fig.1(a) every state is paired randomly with another state at each MC 
step and assumes the probability of that state. Fig.1(b) shows a similar 
simulation but states are placed on a circle and at each MC step a state 
assumes the probability of one of its nearest neighbors chosen randomly. 
The update rule on a circle equalizes the probabilities an order of 
magnitude slower as seen by comparing Fig.1(b) with Fig.1(a). The size 
of ensemble used for Fig.1(b) comprises only 101 states as compared 
with $10^3$ states for Fig.1(a), but it takes nearly $10^4$ iterations 
to reach a fixed state of the ensemble. We also observe an even-odd 
effect in case of Fig.1(b). In this case no fixed states are reached if 
ensemble comprises an even number of states. Fig.1(b) was made only to 
explore a curiosity. Remaining simulations presented in our study are 
based on update rule adopted for Fig1(a). The remarkable property of the 
detailed balance algorithm is that all states in the ensemble eventually 
take the probability of one of the states chosen randomly.

Figure-2 shows a plot of $2 \times 10^3$ probabilities characterizing 
respectively the initial and terminal state of an ensemble of $10^3$ 
states averaged over $10^3$ runs of simulation. Open circles belong to 
initial states and filled circles to terminal states. Initial 
probabilities are drawn from a rectangular distribution over the range 
[0,1]. Thus their value averaged over different runs of simulation is 
expected to be $1/2$. The average probability of each state in the 
initial ensemble is shown by an open circle in Fig.2; it is seen the 
open circles huddle about the value $1/2$ as expected. Detailed balance 
updates terminate when each state in the ensemble acquires the 
probability of one of its initial states chosen randomly. We repeat the 
simulation $10^3$ times obtaining $10^3$ independent terminal 
probabilities. Their average values are shown by filled circles in the 
figure, and are superimposed on the plot of open circles. The 
distribution of filled circles is numerically indistinguishable from the 
input distribution of open circles as maybe expected.

Figure-3 shows small fluctuations in entropy of an ensemble along 
increasing updates of ensemble. As observed in earlier figures, 
approximately 150 MC cycles suffice to bring an initial random state of 
an ensemble to a fixed state where probabilities of all states in the 
ensemble are equal. The figure shows data averaged over $10^3$ runs of 
simulations on an ensemble of $10^3$ states. We show the behavior up to 
250 MC cycles which is well above the time needed for probabilities of 
all states to become equal. Entropy continues to have small fluctuations 
about its average value $ \int_0^1 p \ln{p} dp = 1/4$ over the entire 
period of evolution shown in the figure.

Finally, we show in Figure-4 the spectrum of relaxation times that bring 
all states in an ensemble to have the same probability. The figure shows 
superimposed data of $10^6$ runs of simulations on an ensemble of $10^3$ 
states. The data is normalized such that heights of all data points in 
the figure add up to unity. Notice there is a relatively sharp peak in 
the density of relaxation times.

\section{Discussion}

The origin of Boltzmann's H-theorem lies in his classic work on kinetic 
theory of gases nearly 150 years ago ~\cite{boltzmann}. It is a corner 
stone in the development of statistical mechanics. It has also caused an 
enduring controversy in the field. The theorem is evidently in conflict 
with time reversibility of other fundamental laws of physics. Why 
$H=\sum_i p_i\ln{p_i}$ should only decrease in future if fundamental 
laws of physics do not distinguish between past and future?  The 
question acquires more intrigue in light of ubiquitous human experience 
of reversibility in life. Is the arrow of time something that exists 
solely in human mind? What is time? These remain deep and unresolved 
puzzles even today and therefore research in this field has never 
stopped. It is of course outside the scope of the present note to answer 
these deep questions. We are content to have examined numerically some 
aspects of the theorem that were not clear to us beforehand, and may 
have bearing on the above questions.

We may note that Boltzmann's H-theorem inspired concepts of entropy in 
other fields like quantum mechanics ~\cite{neumann} and information 
theory ~\cite{shannon}. In each case an expression similar to H is a 
measure of uncertainty or disorder in a system. This formula is easy to 
understand intuitively. Events occurring with smaller probabilities 
$p_i$ and correspondingly large values of $\ln{(1/p_i)}$ cause 
relatively large unexpectedness or surprise in the outcome of a physical 
measurement. Thus $H=-\sum_i p_i \ln{(1/p_i)}$ maybe thought as average 
unexpectedness in measurements on a system. Entropy is simply negative of 
H within a multiplicative constant. Neumann's entropy is a measure of 
statistical uncertainty in a quantum description of a system. Indeed, 
the proof $dH/dt \le 0$ for a quantum system closely resembles the proof 
of Boltzmann's theorem that we outlined in the Introduction. Much 
thought has gone into investigating what sneaks into simple and 
transparent derivations that produce irreversibility in both classical 
and quantum cases. In case of dilute classical gas, one generally blames 
an implicit assumption of $\it{stosszahlansatz}$ or molecular chaos. The 
quantum case has no such assumption. It only uses unitarity of time 
evolution operator. The unitarity essentially means the probability of 
transition from a given state to all other states must add up to unity. 
This is most reasonable. However, error is thought to lie in including 
only pure states in the analysis and leaving out mixed states.

Boltzmann also put forth what he called ergodic hypothesis 
~\cite{gallavotti}. It essentially postulates that log-time average of a 
quantity in statistical mechanics is equivalent to its ensemble average. 
The hypothesis is difficult to prove. However its usefulness and value 
is obvious by immense success of statistical mechanics. Equivalence of 
static and dynamic averages should have something to do with the 
H-theorem. Our results demonstrate that detailed balance dynamics 
produces an ensemble of fixed ensembles similar to the initial 
ensemble.The transition time is sharply peaked in thermodynamic limit. 
Sharp peaks in the expectation values of various quantities establish 
broad correspondence of statistical mechanical to that of classic 
thermodynamics based on state functions. Similarly sharp peak in 
relaxation times may indicate why configuration averages reproduce 
observed time averages as well as they do.

}

\begin{figure}[!tbp] 
\centering
\subfloat[]{\includegraphics[width=.35\linewidth,angle=-90]{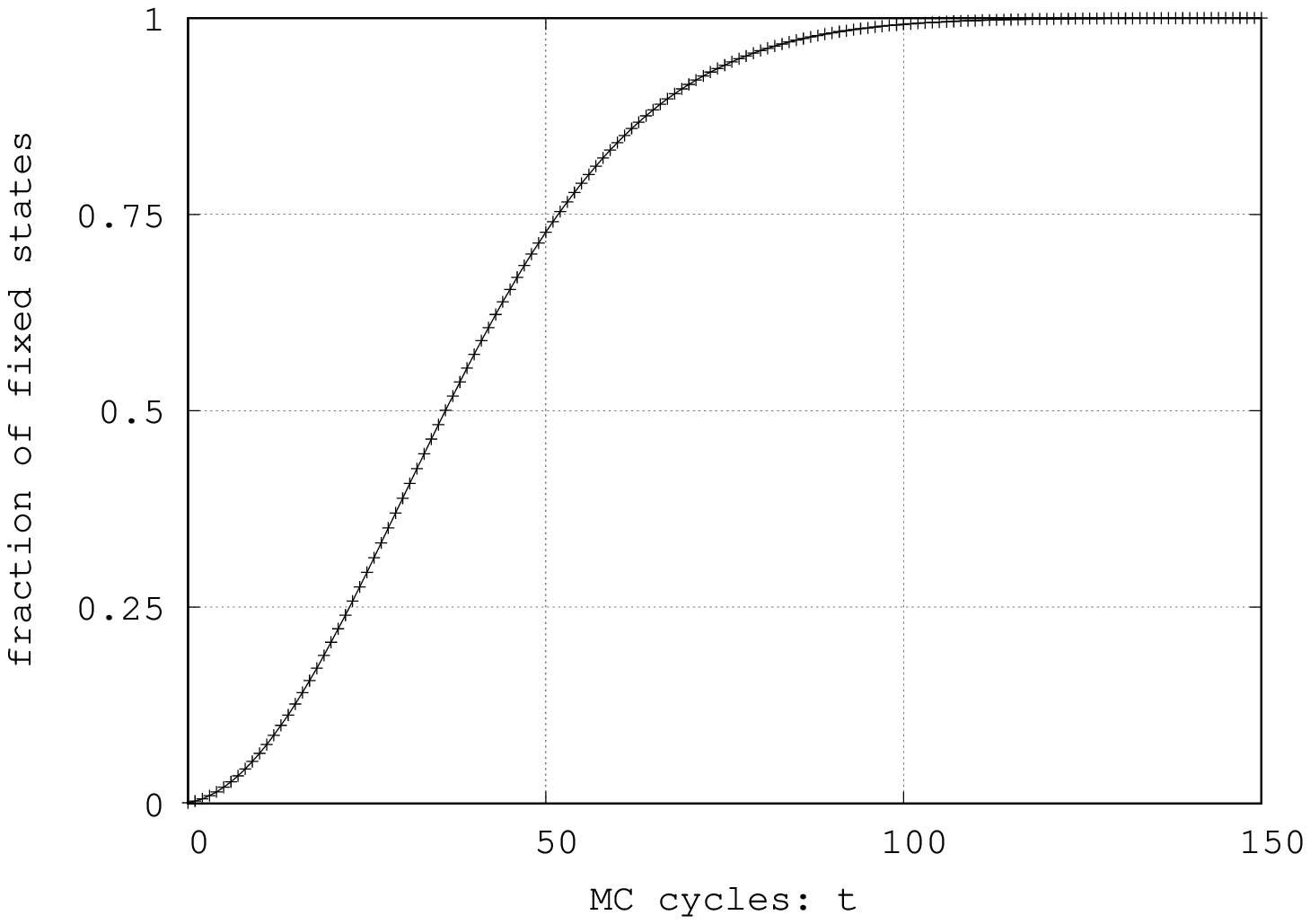}} 
\subfloat[]{\includegraphics[width=.35\linewidth,angle=-90]{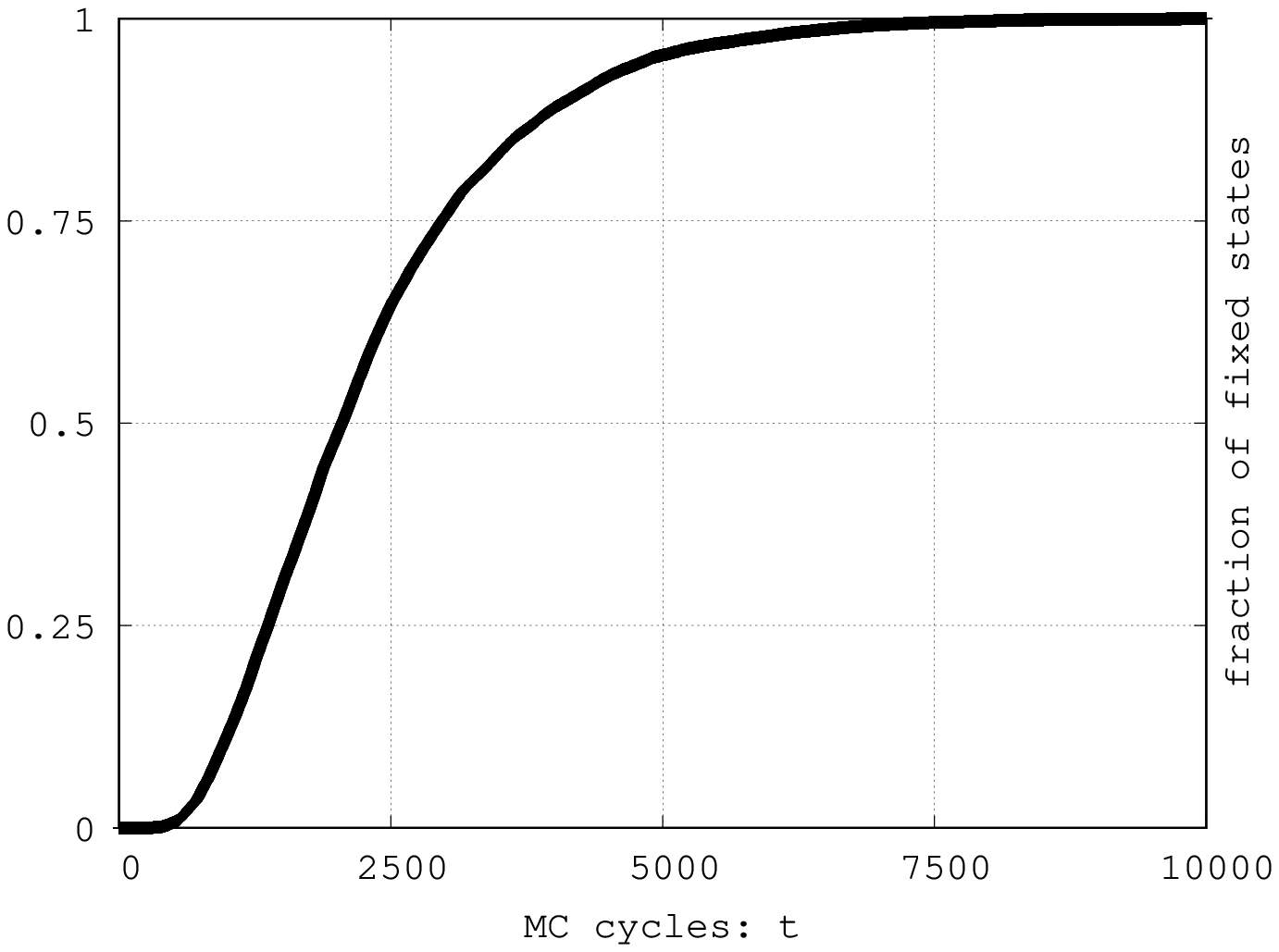}} 

\caption{ Increasing fraction of states in an ensemble that acquire 
equal probability under detailed balance dynamics. Fig.1(a) shows 
simulation on a modest size ensemble of $10^3$ states, initially having 
uniformly distributed probabilities over the interval [0,1]. The x-axis 
shows increasing Monte Carlo steps of parallel dynamics for a randomly 
chosen initial state of an ensemble. The y-axis shows the fraction of 
states in the ensemble that acquire equal probability up to the 
corresponding time on the x-axis. The data is averaged over $10^3$ runs 
of simulation. At each step of a given run, every state is paired 
randomly with another state of the ensemble and assumes the probability 
of that state. In all runs of our simulations, initial states in the 
ensemble acquired equal probability within $150$ iterations. Fig.1(b) 
shows a similar simulation but states are placed on a circle and at each 
step of parallel dynamics, a state assumes the probability of one of its 
nearest neighbors chosen randomly. The update rule on a circle equalizes 
the probabilities much slower as seen by comparing Fig.1(b) with 
Fig.1(a). Note that the data in Fig.1(b) is obtained from a smaller 
ensemble of $10^2+1$ states but it takes nearly $10^4$ iterations in 
each run of simulation for all states to reach equal probability. There 
is an even-odd effect in case of Fig.1(b); ensembles with even number of 
states do not reach a fixed state. Remaining simulations presented in 
our study are based on update rule adopted for Fig1(a).
}
\label{figure:1} \end{figure}

\begin{figure}[!tbp] \centering 
\subfloat[]{\includegraphics[width=.70\linewidth,angle=-90]{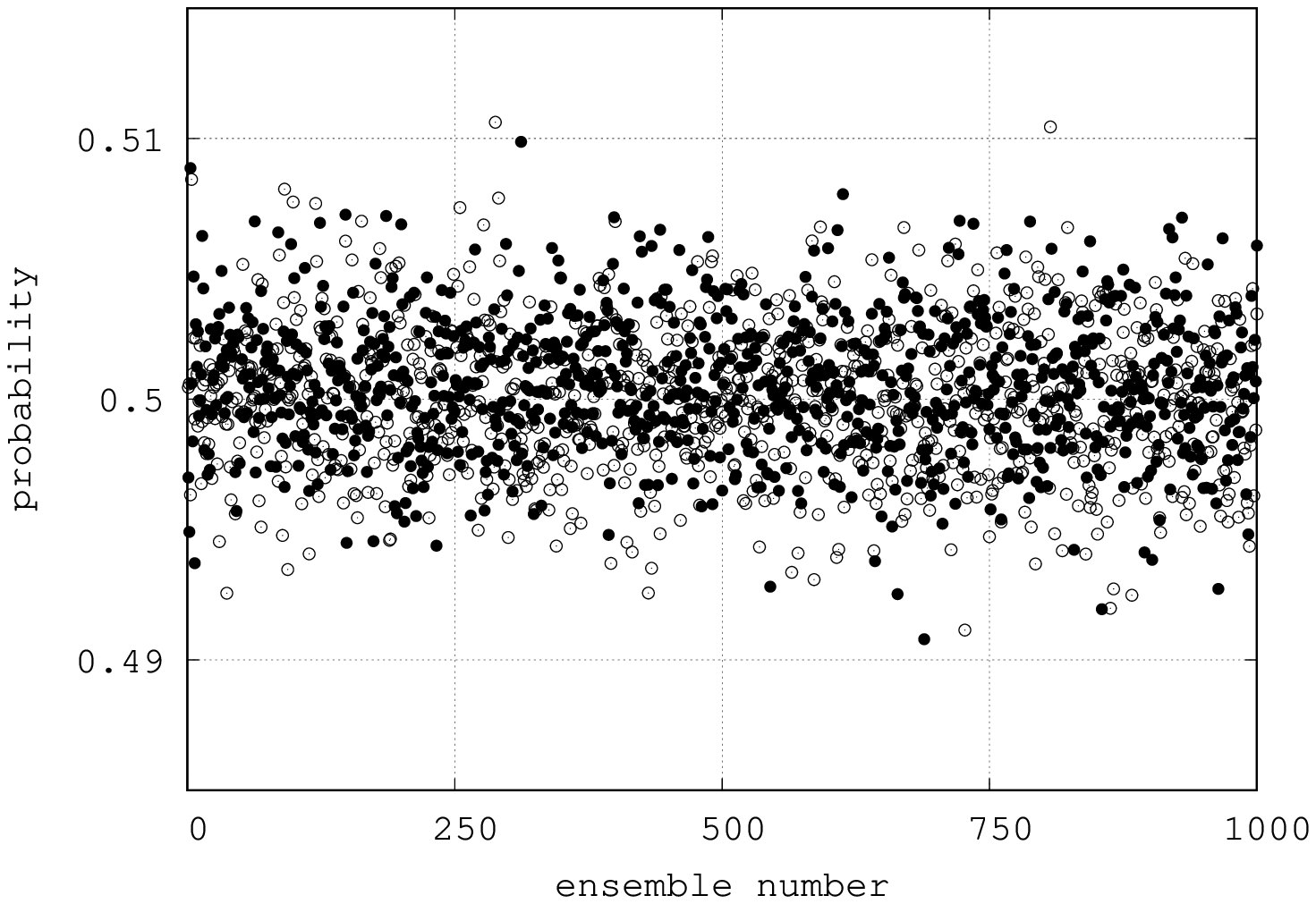}} 
\caption{ Initial and final probabilities of states in an ensemble of 
$10^3$ states under detailed balance dynamics. The initial probabilities 
$\{p_i; i=1, \ldots, 10^3\}$ are chosen from a uniform rectangular 
distribution over the range [0,1].The final probabilities at the end of 
dynamics are constant equal to $p^*$ for every state in the ensemble; 
$p^*$ is equal to one of the initial probabilities $p_i$ chosen randomly. 
The simulation is repeated $10^3$ times. The figure shows average of 
$p_i$ (open circles) and $p^*$ (filled circles) over different runs for 
each state of the ensemble. The distributions $\{p_i\}$ and $\{p^*\}$ 
are numerically indistinguishable, and both are centered around the 
average probability equal to $1/2$ as may be expected.
}
\label{figure:2} \end{figure}

\begin{figure}[!tbp] \centering 
\subfloat[]{\includegraphics[width=.70\linewidth,angle=-90]{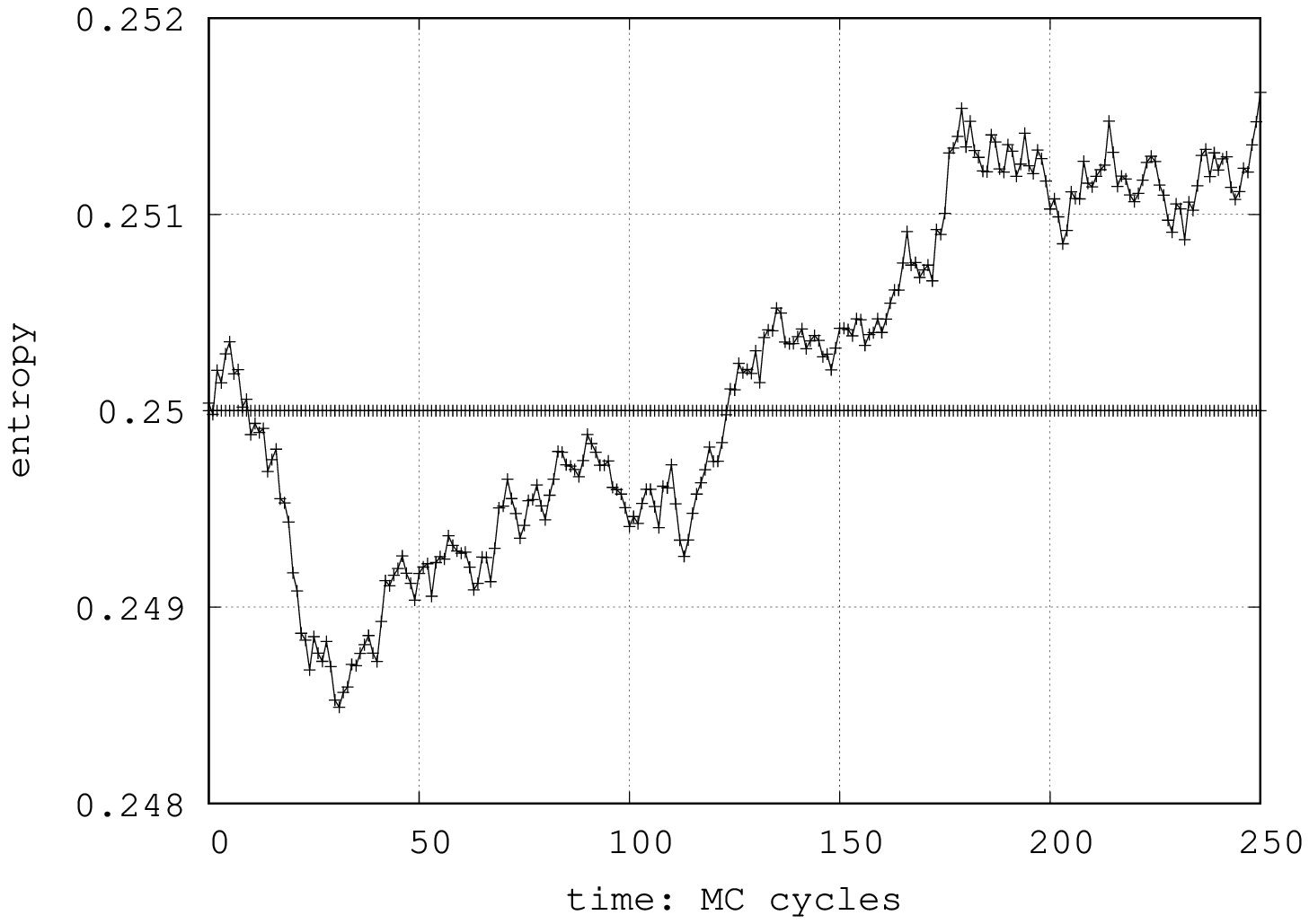}} 
\caption{ This figure presents another aspect of simulations presented 
in earlier figures. It shows entropy of an ensemble at successive steps 
of detailed balance update. Although all states in an ensemble of $10^3$ 
states reach equal probability in about $150$ steps, the value of 
entropy fluctuates about its average value $1/4$ in different runs. Data 
from $10^3$ runs are superimposed.} \label{figure:3} \end{figure}

\begin{figure}[!tbp] \centering 
\subfloat[]{\includegraphics[width=.70\linewidth,angle=-90]{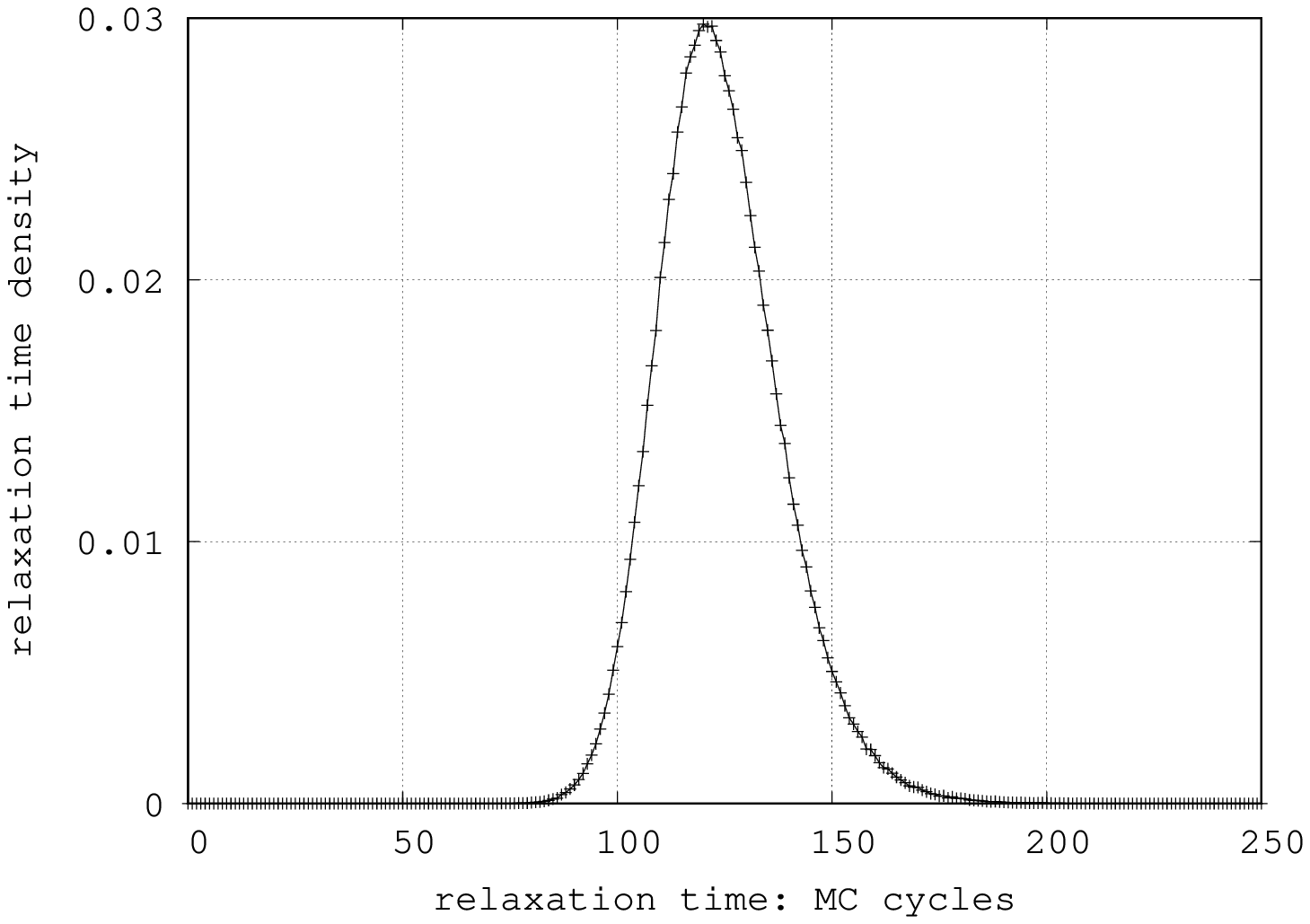}} 
\caption{ Spectrum of relaxation times that bring all states in an 
ensemble to have the same probability. The figure shows superimposed 
data over $10^6$ runs of simulations on an ensemble of $10^3$ states. 
The data is normalized such that heights of all data points in the 
figure add up to unity. Notice there is a relatively sharp peak in the 
density of relaxation times.} \label{figure:4} \end{figure}

}

\end{document}